\documentclass[10pt]{article}
\usepackage{amsmath,amssymb,amsfonts}
\usepackage{units}
\usepackage{bbold,euler}
\usepackage{graphicx}
\usepackage{dcolumn}
\usepackage{bm,bbm,mathtools,esvect}
\usepackage{slashed}
\usepackage{esvect}
\usepackage{fullpage}
\usepackage[utf8]{inputenc}
\title{\bf Generalized imaginary units in quantum mechanics }
\author{{\sf SERGIO GIARDINO\footnote{\tt sergio.giardino@ufrgs.br}}\\
\\
\small \it Departamento de Matem\'atica Pura e Aplicada \\
\small \it Universidade Federal do Rio Grande do Sul (UFRGS)\\
\small \it Caixa Postal 15080, 91501-970  Porto Alegre RS \\
\small \it Brazil}
\begin{document}
\date{} 
\maketitle

\begin{abstract}
\noindent The generalization of the imaginary unit is examined within the instances of the complex quantum mechanics ($\mathbbm C$QM), and of the quaternionic quantum mechanics ($\mathbbm H$QM) as well. Whereas the complex theory  describes non-stationary quantum processes, the quaternionic theory  does not admit such an interpretation, and associates the generalized imaginary unit to a novel time evolution function. Various possibilities are opened as future directions for future research.

\vspace{2mm}

\noindent keywords: quantum mechanics; formalism; other topics in mathematical methods in physics

\vspace{1mm}

\noindent pacs numbers: 03.65.-w; 03.65.Ca; 02.90.+p.

\end{abstract}

\section{ Introduction\label{I}}

In this article, one considers the following question: is it possible to generalize the role played by the imaginary unit in quantum mechanics? Remembering the Schr\"odinger equation,
\begin{equation}\label{i01}
 \hbar \frac{\partial \psi}{\partial t} i=-\frac{\hbar^2}{2m}\nabla^2\psi+V\psi,
\end{equation}
where $\psi=\psi(t,\,\bm x)$ is the complex valued wave function, and $V=V(\bm x)$ is the real valued potential function, we ask whether it is possible to change (\ref{i01}) using a generalized imaginary unit, such as
\begin{equation}\label{i02}
 i\to i(\bm x,\,t).
\end{equation}
One must notice that (\ref{i02}) does not attempt to define a new kind of complex number, but to redefine the imaginary unit required in quantum operators only.  This generalization is motivated by the study of the quaternionic eigenvalue equation \cite{Giardino:2023spz}
\begin{equation}\label{a01}
\frac{d\Lambda}{dx}=\mu\Lambda,
\end{equation}
where $\Lambda$ is a quaternionic function, and $\mu$ is a quaternionic eigenvalue. As will be seen in Section 3 of this paper, these results permit to obtain to obtain novel solutions of time dependent part of the quaternionic Schr\"odinger equation. Although it is natural to hypothesize that (\ref{i02}) can be implemented also in the complex wave equation, the generalized complex unit will be seen to be consistent just in terms of quaternionic quantum mechanics ($\mathbbm H$QM), while the generalization within the usual complex quantum mechanics ($\mathbbm C$QM) can be interpreted either in terms of  in terms of non-stationary processes. 

On the other hand, the quaternionic generalization of the imaginary unit in $\mathbbm H$QM is localized within the scope of quaternionic applications in physics, an important trend in contemporary research. One can mention applications of this sort to the Bell theorem \cite{Gill:2021zmv,Christian:2022leb},
to condensed matter physics \cite{Sbitnev:2023xwr}, to  general relativity \cite{Chanyal:2021gsb,Sbitnev:2019iyz}, to cosmology \cite{Jimenez:2022fll}, and to the
$\mathcal{CPT}$-invariance \cite{Dolan:2020kxj}, among many others. However,  the particular case of quaternionic applications in quantum mechanics depends on the number field that supports the Hilbert space. Quaternionic applications that adopt the Hibert space over complex numbers refer to $\mathbbm C$QM, and can be considered solutions methods \cite{Arbab:2010kr,Soloviev:2016qsx,Sapa:2020dqm,Steinberg:2020xvf,Rawat:2022xsj,Arbab:2022cpe}. Quaternionic quantum applications that cannot be reduced to complex cases involve either a quaternionic Hilbert space, or a real Hilbert space. The $\mathbbm H$QM in the quaternionic Hilbert space is also known as anti-hermitian, because this property is assigned to the Hamiltonian operator in the Schr\"odinger equation. The most important results of anti-Hermitian $\mathbbm H$QM are compiled in a book by Stephen Adler \cite{Adler:1995qqm}. However, this formulation is plagued by an ill-defined classical limit ({\it cf.} Section 4.4 of Adler's book), a drawback that highly reduces the interest to this theory.  A consistent classical limit in $\mathbbm H$QM was achieved by way of the real Hilbert space approach \cite{Giardino:2018lem,Giardino:2018rhs}, what established the foundation for several quaternionic solutions never obtained inside of the anti-Hermitian approach, such as the Aharonov-Bohm effect \cite{Giardino:2016xap}, the free particle \cite{Giardino:2017yke,Giardino:2017pqq}, the Virial theorem\cite{Giardino:2019xwm}, the quantum scattering \cite{Giardino:2020ztf,Hasan:2019ipt}, the rectangular potentials \cite{Giardino:2020cee}, and the harmonic oscillator \cite{Giardino:2021ofo}. Relativistic quaternionic solutions were also obtained through the real way, namely the
quaternionic Klein-Gordon \cite{Giardino:2021lov}, the Dirac equation \cite{Giardino:2021mjj}, the scalar field \cite{Giardino:2022kxk}, and the fermionic field \cite{Giardino:2022gqn}. The real way to $\mathbbm H$QM will thus be considered to examine the generalization of the imaginary unit, and the results will be contrasted with the complex case.

As a final comment, one may wonder the relation between $\mathbbm H$QM in the real Hilbert space, and real quantum mechanics ($\mathbbm R$QM), a subject of contemporary interest, remembering that $\mathbbm R$QM was originally proposed by Ernst Stueckelberg \cite{Stueckelberg:1960rsi,Stueckelberg:1961rsi,Stueckelberg:1961psg,Stueckelberg:1962fra}, and this theory was considered equivalent to $\mathbbm C$QM. A recent controversy in the literature appeared because of results that could refute $\mathbbm R$QM as a viable quantum theory  \cite{Renou:2021dvp,Chen:2021ril,Wu:2022vvi}, while other researchers consider that this is not the case \cite{Finkelstei:2022rqm,Chiribella:2022dgr,Zhu:2020iml,Fuchs:2022rih,Vedral:2023pij}. Independent of the answer that will emerge, this discussion does not concern to the subject of the present article, because $\mathbbm H$QM in the Hilbert space use quaternionic states, non-Hermitian operators and a specific real valued inner product, and hence is clearly not equivalent neither to $\mathbbm R$QM, nor to  $\mathbbm C$QM.   In summary, $\mathbbm H$QM in the real Hilbert space is a theory whose defining features are still unknown, and this article presents a further aspect of their difference to the current and accepted theories of quantum mechanics.

In summary, the consideration of the proposal (\ref{i02}) deepens the comprehension of the role of the complex unit in quantum mechnics, as well as sheds additional light on the differences between $\mathbbm H$QM, and $\mathbbm C$QM reinforcing that they are different theories that cannot be totally     identified. Moreover, the hypothesis of the quaternionic version as the most general theory one is also accordingly fortified.

\section{ The complex case \label{CC}}
Considering the complex instance, which is simpler, and will be held as a reference to the quaternionic one, the hypothesis (\ref{i02}) is introduced in (\ref{i01})  as
\begin{equation}\label{cc01}
 \hbar \frac{\partial \psi}{\partial t} e^{i\theta}i=\frac{1}{2m}\widehat{p}^2\psi+V\,\psi,
\end{equation} 
where $\widehat{\bm p}=-i\hbar\bm\nabla$ is the momentum operator, and $\theta=\theta(\bm x,\,t)$ is a real function. Equation (\ref{cc01}) can be understood by means of the transformation
\begin{equation}
 \varphi=e^{i\theta}\psi
\end{equation}
which implies 
\begin{equation}\label{cc15}
 \hbar \frac{\partial \varphi}{\partial t} e^{i\theta}i=\frac{1}{2m}\widehat{\pi}^2\varphi+\mathcal V\,\varphi.
\end{equation}
The generalized momentum $\widehat\pi$, and the generalized scalar potential $\mathcal V$ respectively reads
\begin{equation}
 \widehat{\bm\pi}=\widehat{\bm p}-\hbar\,\bm A,\qquad\mathcal V=V-\hbar e^{i\theta}\frac{\partial \theta}{\partial t}
\end{equation}
and the gauge vector potential $\bm{\mathcal A}$ accordingly is
\begin{equation}
 \bm{\mathcal A}=\bm\nabla\theta.
\end{equation}
 Therefore, equations (\ref{cc01}) and (\ref{cc15}) are identical, indicating that $i\to i\exp[i\theta]$ is not a gauge transformation that can be absorbed in a wave function of a stationary quantum state. The physical meaning of this transformation is obtained from the deformed continuity equation obtained from (\ref{cc01}), namely
\begin{equation}\label{cc16}
 \cos\theta\frac{\partial \rho}{\partial t}+\bm{\nabla\cdot j}=\kappa+\lambda,
\end{equation}
where the probability density $\rho$ and the current density $\bm j$ are the familiar
\begin{equation}
 \rho=\psi\psi^\dagger, 
\qquad\mbox{ and }\qquad
\bm j=\frac{1}{2m}\Bigg[\Big(\widehat{\bm p} \psi\Big) \psi^\dagger+\Big({\widehat{\bm p} \psi}\Big)^\dagger \psi\, \Bigg],
\end{equation}
where $\psi^\dagger$ denotes the hermitian conjugate of $\psi$. The $\kappa$ and $\lambda$ terms on the right hand side of (\ref{cc16}) indicate that the probability is not conserved. Their expressions are 
\begin{equation}
 \kappa=i\sin\theta\left(\psi\frac{\partial\psi^\dagger}{\partial t}-\psi^\dagger\frac{\partial\psi}{\partial t}\right)
 \qquad\mbox{ and }\qquad
 \lambda=\frac{i}{\hbar}\rho\big(V^\dagger-V\big).
\end{equation}
The $\lambda$ contribution disappears in the case of a real potential $V$, as expected, while $\kappa$ disappears if $\theta=0$, also as desired. If $\theta$ is small, we approximately obtain the continuity equation with a non-zero contribution due to $\theta$ to the non-conservative terms on the right hand side of (\ref{cc16}). In summary, we observe that the transformation $i\to i\exp[i\theta]$ is associated  with non-stationary quantum processes, usually described in terms of complex potentials ({\it cf.} Section 20 of \cite{Schiff:1968qmq}).

A slightly different solution for (\ref{cc01}) can be obtained via
\begin{equation}\label{cc02}
 \psi=\phi(x) \exp\left[-\frac{i}{\hbar}Et(\cos\theta-i\sin\theta)\right],
\end{equation}
where $E$ is, of course, the real energy, and $\phi(x)$ is a complex valued function. The usual eigen-value equation
\begin{equation}\label{cc03}
 -\frac{\hbar^2}{2m}\nabla^2\psi+V\psi=E\psi
\end{equation}
is obtained, and $\,\phi\,$ satisfies (\ref{cc03}) if $\,\theta=\theta(t)\,$, as expected.  Moreover, the probability density satisfies the continuity equation 
\begin{equation}\label{cc04}
 \frac{\partial \rho}{\partial t}+\bm{\nabla\cdot J}=\beta+\gamma,
\end{equation}
and the presence of the source terms $\beta$ and $\gamma$ immediately notices the probability current not to be conserved. The terms of (\ref{cc04}) are as follows: the probability density is the familiar
\begin{equation}\label{cc05}
 \rho=\psi\psi^\dagger, 
\end{equation}
 However, the current density $\bm J$ differs from the common definition. Accordingly,
\begin{equation}\label{cc06}
 \bm J=\frac{1}{2m}\Bigg[\Big(\widehat{\bm p} \psi\Big) \psi^\dagger e^{-i\theta}+\Big({\widehat{\bm p} \psi}\Big)^\dagger \psi\, e^{i\theta}\Bigg],
\end{equation}
and setting  $\theta=0$ recovers the conventional result, as wished. The first source term reads
\begin{equation}\label{cc07}
 \beta=\frac{1}{\hbar}\,i\,\rho\Big( V^\dagger e^{i\theta}-V e^{-i\theta}\Big),
\end{equation}
and it recovers the usual source term is recovered if $\theta=0$, as expected. However, in the case of a real potential, namely $V^\dagger=V$, 
\begin{equation}\label{cc08}
 \beta=-\frac{2 V}{\hbar}\rho\,\sin\theta,
\end{equation}
and then a real potential may also have a non-conservative probability if (\ref{cc01}) and (\ref{cc02}) hold.  Non-conservative probabilities are usually found in quantum processes described in terms of complex potentials \cite{Schiff:1968qmq}, and their expression in terms of real potentials is certainly a novel and remarkable property.  Finally, the last source term of (\ref{cc04}) is as follows
\begin{equation}\label{cc09}
 \gamma=J_0-\frac{1}{m\hbar}\big|\widehat{\bm p}\psi\big|^2\sin\theta,
\end{equation}
where
\begin{equation}\label{cc10}
 J_0=\frac{1}{2m\hbar}\Bigg[\big(\widehat{\bm p} \, \psi\bm\cdot\widehat{\bm p}\theta\big)\psi^\dagger \,e^{-i\theta}+\big(\widehat{\bm p} \psi\bm\cdot\widehat{\bm p}\theta\big)^\dagger\, \psi\, e^{i\theta}\Bigg].
\end{equation}
The usual $\mathbbm C$QM result is recovered for constant $\theta$, however, (\ref{cc08}-\ref{cc09}) permit one to interpret $\theta$ as a parameter, or even as a generator for non-conservative quantum processes. This simple fact indicates a way to relate conservative and non-conservative quantum processes, and eventually Hermitian, and non Hermitian quantum physics. 

A novel imaginary constant could be defined satisfying (\ref{i02}). Expressly, 
\begin{equation}\label{cc11}
 \eta=e^{i\theta}i,
\end{equation}
but does not satisfy the necessary property
\begin{equation}\label{cc12}
 \eta^2=-1,
\end{equation}
although it satisfies
\begin{equation}
 \eta\bar\eta=1,
\end{equation}
where $\bar\eta$ is the complex conjugate of $\eta$. Therefore, $\eta$ does not generalize $i$, and it cannot consistently redefine the quantum operators. By way of example, the density current (\ref{cc06}) inspires the proposition of the generalized momentum
\begin{equation}\label{cc13}
\widehat{ \bm\varpi}=-\hbar\, e^{-i\theta}\,\bm\nabla,
\end{equation}
but
\begin{equation}
 \widehat{ \bm\varpi}^2=\hbar^2\, e^{-i2\theta}\,\nabla^2
\end{equation}
and therefore this operator does not recover the Hamiltonian operator for the Schr\"odinger equation for arbitrary $\theta$. Therefore,  (\ref{cc01}) may be considered as a particular deformation of the Schr\"odinger equation, but it does not  generalize the whole theory.

As a final comment, one can consider the commutation relation along the $x$ direction
\begin{equation}\label{cc14}
 \big[\,\widehat x,\,\widehat\varpi_x\big]=\hbar\, i\,e^{i\theta}.
\end{equation}
Despite the inconsistency of the momentum operator $\,\widehat\varpi_x\,$, this commutator permits to suppose a general uncertainty principle connected to non-conservative processes, where particles can be either created, or destroyed. At least, in principle. It can be seen that (\ref{cc14}) admits a real term in the right hand side, indicating that the imaginary term is associated to the stationary processes, and the real term is associated to decaying processes. This observation is not considered in usual theories of the Generalized Uncertainty Principle (GUP) \cite{Tawfik:2015rva}, and is an interesting possibility for future research. 

In the next section, the ideas schematically hypothesized in the complex case will be consistently built within a quaternionic framework.

\section{The quaternionic case}

This article is not intended to introduce the mathematical foundations of quaternions. Among a huge amount of references, one can mention \cite{Morais:2014rqc,Garling:2011zz,Ward:1997qcn} as good introducing texts, as well as the introductory section of \cite{Giardino:2021onv} for many of the definitions used in this article. Let us entertain  the unitary quaternion
\begin{equation}\label{gi01}
 \Lambda=\cos\Theta \,e^{i\Gamma}+\sin\Theta \,e^{i\Omega}\,j,
\end{equation}
where 
\begin{equation}\label{gi012}
\Theta=\Theta(\bm x,\,t),\qquad\Gamma=\Gamma(\bm x,\,t),\qquad\mbox{ and}\qquad \Omega=\Omega(\bm x,\,t)
\end{equation}
are differentiable real functions. In a recent paper \cite{Giardino:2023spz}, it has been observed  that the unit quaternion (\ref{gi01}) satisfies 
\begin{equation}\label{gi02}
 \frac{d\Lambda}{dt}=\frac{d\Theta}{dt}\,\Lambda\,\eta, 
\end{equation}
wherever $\Gamma=\Gamma(\bm x),$  $\Omega=\Omega(\bm x)$, 
\begin{equation}\label{gi03}
 \eta=e^{i\Xi}j,\qquad\mbox{and}\qquad \Xi=\Omega-\Gamma.
\end{equation}
Additionally,
\begin{equation}\label{gi04}
 \eta^2=-1,
\end{equation}
remembering the necessary property (\ref{gi04}) that does not hold in the generalization (\ref{cc11}).
The general quaternionic unit $\eta$ permits a novel definition for the wave equation, whose quaternionic wave function reads
\begin{equation}\label{gi044}
 \Psi=\Psi_0+\Psi_1 j,
\end{equation}
with $\Psi_0$, and $\Psi_1$ complex functions. In order to be a solution to a wave equation, such a wave function can be decomposed as
\begin{equation}\label{gi044a}
\Psi(\bm x,\,t)=\Phi(\bm x)\Lambda(\bm x,\,t),
\end{equation}
where $\Phi$ is also quaternionic. Therefore, the generalized Schr\"odinger equation is proposed to be
\begin{equation}\label{gi05}
 \hbar\frac{\partial\Psi}{\partial t}\eta=\widehat{\mathcal H}\Psi,
\end{equation}
remembering the mandatory multiplication of $\eta$ on the right hand side of the time derivative of the wave function, as defined in the real Hilbert space formulation of $\mathbbm H$QM \cite{Giardino:2018lem,Giardino:2018rhs}.  Using (\ref{gi02}) and  (\ref{gi044a}) in (\ref{gi05}), it is obtained that
\begin{equation}\label{sch}
 -\hbar\frac{\partial\Theta}{\partial t} \Psi=\widehat{\mathcal H}\Psi.
\end{equation}
In the case where $\Lambda$ does not depend on $x$, and $\Theta$ is a linear real function of the time variable, (\ref{sch}) recovers the time independent Schroedinger equation. Therefore, the time dependent function $\Lambda$ can be eliminated, and (\ref{sch}) becomes an eigenvalue equation in terms of the time independent quaternionic wave function $\Phi$.

Also within the real Hilbert space approach,   the  quaternionic Hamiltonian operator states
\begin{equation}\label{gi06}
\mathcal H=-\frac{\hbar^2}{2m}\Big(\bm\nabla- \bm{\mathcal A}\Big)^2+U,
\end{equation}
in which the quaternionic vector potential $\bm{\mathcal A}$, and the quaternionic scalar potential $U$ are respectively decomposed as
\begin{equation}\label{gi07}
\bm{\mathcal A}=\bm\alpha i+\bm\beta j,\qquad\mbox{and}\qquad U=V +W\,j.
\end{equation}
Accordingly, $\bm\alpha$ is a real vector function, $\bm\beta$ is a complex vector function and $V$ and $W$ 
are complex scalar functions. In order to verify whether (\ref{gi05}) satisfies the necessary consistency conditions required to generate trustful 
quantum solutions, one observes  that the continuity equation similar to (\ref{cc04})
\begin{equation}\label{gi08}
 \frac{\partial \mathcal P}{\partial t}+\bm{\nabla\cdot \mathcal J}=\mathcal B+\mathcal G,
\end{equation}
where the probability density, as expected, is
\begin{equation}\label{gi09}
 \mathcal P=\Psi\,\Psi^\dagger,
\end{equation}
the probability current is
\begin{equation}\label{gi10}
\bm{\mathcal J}=\frac{1}{2m}\left[\big(\widehat{\bm \Pi} \Psi\big)\, \Psi^\dagger \,+\big(\widehat{\bm \Pi} \Psi\big)^\dagger\, \Psi\,\right],
\end{equation}
and the generalized momentum operator accordingly is
\begin{equation}\label{gi11}
 \widehat{\bm\Pi}\Psi=-\hbar\Big[\big(\bm\nabla-\bm{\mathcal{A}}\big)\Psi\Big]\eta.
\end{equation}
The generalization of the usual momentum obtained by $i\to\eta$ is consistent, and the kinectic term of the Hamiltonian operator as a squared momentum
also holds. The non-conservative source terms of the probability density are as follows
\begin{equation}\label{gi12}
 \mathcal B=\frac{1}{\hbar}\Big(\,U\Psi\,\eta\,\Psi^\dagger-\Psi\,\eta\,\Psi^\dagger U^\dagger\,\Big),
\end{equation}
and
\begin{equation}\label{gi13}
 \mathcal G=\frac{1}{2m\hbar}\left[\Big(\widehat{\bm\Pi}\Psi\bm{\cdot}\widehat{\bm p}\Xi\Big)\Psi^\dagger+
 \Psi\,\Big(\widehat{\bm p}\Xi\Big)^\dagger\bm\cdot\Big(\widehat{\bm\Pi}\Psi\Big)^\dagger\,\right].
\end{equation}
The source term (\ref{gi12}) goes to zero in case of a real scalar potential $U$, what conforms to the theory of complex potentials in $\mathbb C$QM  \cite{Schiff:1968qmq}, but disagrees to the source term $\beta$ of (\ref{cc07}-\ref{cc08}), typifying $\eta$ as a generalization of the complex unit. On the other hand, the source terms $\gamma$ (\ref{cc09}), and $\mathcal G$ (\ref{gi13}), differ because of the absence of a correspondent in $\mathcal G$ to the second term of $\gamma$. However, the first term of $\gamma$, namely $J_0$, conforms exactly to $\mathcal G$,  meaning that a general imaginary unit that is  also a function of the space coordinates induces a source  to the probability current. 

In summary, the imaginary unit $\eta$ defined in (\ref{gi03}) generalizes the imaginary unit $i$, it enables consistent the quantum operators, and it can be turned into a function, what means to be a local symmetry, or a gauge symmetry. The general imaginary unit also holds within the uncertainty relation
\begin{equation}\label{gi14}
 \big[x,\,\Pi_x\big]=(\hbar|\eta),
\end{equation}
where $(a|b)$ impose an order of multiplication factor, namely $(a|b)c=acb$. The commutator (\ref{gi14}) further confirms the general character of $\eta$. As a final comment, the general Schr\"odinger equation (\ref{gi05}) also holds if the pure spatially dependent term of the wave funtion (\ref{gi044a}), accordingly $\Phi(\bm x)$, is a complex function. This is another remarkable fact, and raises up the  possibility of a wave function where only the time evolution  associated to $\eta$ is quaternionic, and the space dependent part is complex. An investigation concerning the differences between the complex and the quaternionic time evolution is of course another very interesting direction for future research.

\section{Complex solutions\label{D}}
This section aims to be a reference to the next quaternionic section, in the same token as Section 2 is a reference to Section 3. In order to entertain the solutions for (\ref{cc01}), one proposes the wave function
 \begin{equation}\label{d001}
  \psi(\bm x,\,t)=\chi(\bm x,\,t)\,\phi(\bm x),\qquad\mbox{where}\qquad\chi(\bm x,\,t)=\exp\Big[-i\,\mathcal E e^{-i\theta}\Big],
 \end{equation}
and with the real functions $\mathcal E=\mathcal E(t)$, and $\theta=\theta(\bm x,\,t)$. Of course, the usual quantum wave function is recovered in case of
\begin{equation}\label{d002}
 \mathcal E=\frac{E}{\hbar} t,\qquad\qquad \theta=0,
\end{equation}
and $\,E\,$ a real constant interpreted as the stationary energy. Therefore, (\ref{cc01}) and (\ref{d001}) give
\begin{equation}\label{d003}
 \hbar\Big(\dot{\mathcal E} -i\dot\theta\mathcal E\Big)\phi=-\frac{\hbar^2}{2m}\Bigg[\nabla^2\phi+\Big(i\phi|\bm\nabla\theta|^2-\phi\nabla^2\theta-2\bm\nabla\phi\bm{\cdot\nabla}\theta\Big)\mathcal E e^{-i\theta}+\phi|\bm\nabla\theta|^2\mathcal E^2e^{-2i\theta}\Bigg]+V\phi,
\end{equation}
where the upper dot sign means a time derivative. In principle, one have to make $\phi(\bm x)\to\phi(\bm x,\,t)$, and expand $\phi$ as a complex Fourier series in order to study more general functions. However, we postpone this more general study to be approached in a separate paper, and focus the simpler situations. Determining,
\begin{equation}
\bm\nabla\theta=\bm 0,\qquad\bm \dot\theta=0,\qquad \mathcal E=\frac{E}{\hbar}t,
\end{equation}
 the right hand side of (\ref{d003}) becomes equivalent to what is found in the time independent Schr\"odinger equation. However, the left hand side presents additional possibilities. The simplest case, where $\dot\theta=0$, recovers the usual Schr\"odiner equation, but the system remains non-conservative in terms of the probability density because of the non-zero $\beta$ term (\ref{cc08}), what can be seen also from the decaying amplitude of the wave function
\begin{equation}
 |\psi|=|\phi|e^{-\sin\theta\frac{E}{\hbar}t}.
\end{equation}
Therefore, an stationary and normalizable time independent wave function $\phi$ turns either evanescent after choosing $\theta\in(0,\,\pi)$, or increasing whether $\theta\in(\pi,\,2\pi)$. The amplitude of the time dependent wave function is the factor that will determine the conservation of probability, and one can interpret this factor as associated either to creation, or annihilation of particles.
A second possible solution reads
\begin{equation}
 \bm\nabla\theta=\bm 0,\qquad\bm \theta=\theta_0 t,\qquad \dot{\mathcal E}=0,
\end{equation}
where $\theta_0$ is obviously a real constant. This situation is interesting because it admits pure imaginary eigen-values, what are interesting to non-hermitian studies \cite{amore:2007cev,Ashida:2020dkc}. Finally, the most general case is such as
\begin{equation}
 \dot{\mathcal E} -i\dot\theta\mathcal E=\frac{\epsilon}{\hbar}e^{i\xi},
\end{equation}
where $\epsilon$ and $\xi$ are real constants. In this case,
\begin{equation}
 \bm\nabla\theta=\bm 0,\qquad\bm \theta=\theta_0+\frac{\hbar}{E}\sqrt{1-\left(\frac{E}{\epsilon}\right)^2}\ln t,\qquad \dot{\mathcal E}=\frac{E}{\hbar}t
\end{equation}
This last solution indicates the wave function (\ref{cc02}) as either non-conservative probabilities, or non-hermitian processes. These are evidently exciting directions for future investigation, particularly their relation to gauge transformations and geometric phases. Furthermore, these are enough examples that will work as references to the wave functions of the generalized imaginary unit considered in the next section.

\section{Quaternionic solutions\label{Q}}
Let us propose the quaternionic wave function
\begin{equation}\label{q001}
 \Psi(\bm x,\,t)=\Phi(\bm x)\Lambda(\bm x,\,t),
\end{equation}
where $\Phi$ is a quaternionic function, and $\Lambda$ was defined in (\ref{gi01}-\ref{gi012}). Therefore,
\begin{equation}\label{q002}
 \dot\Lambda=\dot\Theta\Lambda\eta+i\Big(\dot\Omega\,\sin\Theta\, e^{i\Gamma}-\dot\Gamma\cos\Theta \,e^{i\Omega}j\Big)\eta,
\end{equation}
and then
\begin{equation}\label{q003}
 \dot\Lambda\,\eta\,\overline\Lambda=-\,\dot\Theta + i\Bigg[\Big(\dot\Gamma-\dot\Omega\Big)\sin\Theta\cos\Theta+\Big(\dot\Gamma\cos^2\Theta+\dot\Omega\sin^2\Theta\Big)e^{i(\Gamma+\Omega)}\,j\,\Bigg].
\end{equation}
In order to separate the variables, 
\begin{equation}\label{q004}
 \Theta=\frac{E}{\hbar}t.
\end{equation}
Of course, the whole expression on the right hand side will be a real constant if
\begin{equation}\label{q005}
 \dot\Gamma=\dot\Omega=0.
\end{equation}
However, this is not the unique possibility. The pure imaginary quaternionic expression is eliminated if
\begin{equation}\label{q006}
 \dot\Gamma=F\,\sin^2\Theta\qquad \dot\Omega=-F\,\cos^2\Theta
\end{equation}
where $F=F(\bm x,\,t)$ is a real function. If 
\begin{equation}\label{q007}
F\propto\sec\Theta \csc\Theta,
\end{equation}
then an imaginary constant will appear, and $\Gamma$, and $\Omega$ will be singular. These solutions are be interesting for finite time processes, like decaying, and emission processes, what conforms the usual interpretation for complex eigenvalues of the Schr\"odinger equation. Independent of this, the simplest solution, where (\ref{q004}-\ref{q005}) hold, is therefore
\begin{equation}\label{q008}
 \Lambda=\cos\left(\frac{E}{\hbar}t\right) e^{i\Gamma_0}+\sin\left(\frac{E}{\hbar}t\right) e^{i\Omega_0}j,
\end{equation}
with constant $\Gamma_0$, and $\Omega_0$. The time evolution generated by (\ref{q008}) is novel in $\mathbbm H$QM because the known solutions are the usual complex exponential $e^{-i\Theta}$ usually obtained in $\mathbbm C$QM. The consequences of this solution, if any, have to be understood in future research. 
 
 As a final aspect of the solution to be considered, one can discuss the case where $\Lambda$ contains a space dependence. In this situation,
\begin{equation}\label{q009}
\hbar\Phi\dot\Lambda\eta=-\frac{\hbar^2}{2m}\Big(\nabla^2\Phi\Lambda+2\bm\nabla\Phi\bm{\cdot\nabla}\Lambda+\Phi\nabla^2\Lambda\Big)+V\Phi\Lambda.
\end{equation}
Evidently, the space independent situation, where $\bm\nabla\Lambda=\bm 0$ recovers the usual case. In order to establish a more general discussion, one have to point out that
\begin{equation}\label{q010}
 \bm\nabla\Lambda=\bm P\,e^{i\Gamma}+\bm Q\,e^{i\Omega}\,j
\end{equation}
where
\begin{equation}\label{q011}
 \bm P=-\sin\Theta\,\bm\nabla\Theta+i\cos\Theta\bm\nabla\Gamma,\qquad\mbox{and}\qquad \bm Q=\cos\Theta\,\bm\nabla\Theta+i\sin\Theta\bm\nabla\Omega.
\end{equation}
A simple space-dependent $\Lambda$ is obtained whether the spatial dependence of $\Theta,\,\Gamma$, and $\Omega$ is orthogonal to the spatial dependence in $\Phi$, and evidently $\bm\nabla\Phi\bm{\cdot\nabla}\Lambda=0$. Moreover,
\begin{equation}\label{q012}
 \nabla^2\Lambda=\mathcal M\,e^{i\Gamma}+ \mathcal N\,e^{i\Omega}\,j
\end{equation}
where
\begin{equation}\label{q013}
\mathcal M=-\Big(\big|\bm\nabla\Theta\big|^2+\big|\bm\nabla\Gamma\big|^2\Big)\cos\Theta-\sin\Theta\,\nabla^2\Theta+i\Big(\cos\Theta\,\nabla^2\Gamma
-2\sin\Theta\,\bm\nabla\Theta\bm{\cdot\nabla}\Gamma\Big)
\end{equation}
and
\begin{equation}\label{q014}
\mathcal N=-\Big(\big|\bm\nabla\Theta\big|^2+\big|\bm\nabla\Omega\big|^2\Big)\sin\Theta+\cos\Theta\,\nabla^2\Theta+i\Big(\sin\Theta\,\nabla^2\Omega
+2\cos\Theta\,\bm\nabla\Theta\bm{\cdot\nabla}\Omega\Big).
\end{equation}
Proposing the eigenvalue relation
\begin{equation}\label{p015}
 \nabla^2\Lambda=-\mathcal K\Lambda,
\end{equation}
where $\mathcal K$ is a real constant. After choosing
\begin{equation}\label{q016}
\nabla^2\Theta=\nabla^2\Gamma=\nabla^2\Omega=0,\qquad\bm\nabla\Theta\bm{\cdot\nabla}\Gamma= \bm\nabla\Theta\bm{\cdot\nabla}\Omega=0,
\end{equation}
and also
\begin{equation}\label{q017}
 \bm\nabla\Gamma=\bm\nabla\Omega,
\end{equation}
thus (\ref{q009}) becomes the well known eigenvalue equation
\begin{equation}\label{q018}
 -\frac{\hbar^2}{2m}\nabla^2\Phi+V\Phi=\big(\mathcal E+\mathcal K\big)\Phi.
\end{equation}
That the complex case only supports evanescent solutions, and contrasts the quaternionic that case supports evanescent solutions in finite time processes (\ref{q007}), as well as stationary solutions from (\ref{q018}). This observation is the final element to conclude that the quaternionic imaginary unit $\eta$ defined in (\ref{gi03}) in fact determines a generalization of the imaginary unit $i$ in quantum mechanics, something that is not possible using $e^{i\theta}i$.

\section{Conclusion\label{C}}

This paper responds affirmatively to the question whether it is possible to generalize the imaginary unit $i$ that appears in the Schr\"odinger equation. However, a purely complex generalization such as $e^{i\theta}i$ does not achieve this target, and is rather related to evanescent solutions, where physical processes involving non-conservative probabilities take place. On the other hand, the quaternionic imaginary unit $\eta$, defined in (\ref{gi03}), admits the conventional stationary state wave functions, and provides an authentic generalization of the imaginary unit in the Sch\"odiger equation, and in the quantum operators as well. 

The consequences of this more general framework are unknown for the time being, although it suggests exciting directions for future research. Among others, one can mention the understanding of the time evolution whose simplest possibility was determined in (\ref{q008}), and also the investigation of Fourier series solutions for  (\ref{d003}), the generalized momentum (\ref{cc13}), and the Generalized Uncertainty Principle. Non-hermitian processes are important as well, particularly because of the presence of imaginary eigenvalues, and dissipative processes. In summary, there are several possibilities to be exploited after the determination of a generalized imaginary unit, whose simple existece has the faculty to generalize the wave equation, and possibly their solutions and interpretations.

\bigskip


\paragraph{Data availability statement} The author declares that data sharing is not applicable to this article as no data sets were generated or analysed during the current study.

\paragraph{Declaration of interest statement} The author declares that he has no known competing financial interests or personal relationships that
could have appeared to influence the work reported in this paper.

%
%
%
%
\begin{footnotesize}
\bibliographystyle{unsrt} 
\bibliography{bib_imag}
\end{footnotesize}
\end{document}